\documentclass[a4paper,12pt]{article}
\usepackage{amsthm,amsfonts,amsmath,amssymb,mathrsfs,url}

\title{Bohmian Trajectories as the Foundation\\ of Quantum Mechanics}

\author{
Sheldon Goldstein\footnote{Departments of Mathematics, Physics and
     Philosophy, Rutgers University, 
     Hill Center, 110 Frelinghuysen Road, Piscataway, NJ 08854-8019, USA.
     E-mail: oldstein@math.rutgers.edu},
Roderich Tumulka\footnote{Department of Mathematics,
     Rutgers University, Hill Center,  
     110 Frelinghuysen Road, Piscataway, NJ 08854-8019, USA.
     E-mail: tumulka@math.rutgers.edu},
 and Nino Zangh\`\i\footnote{Dipartimento di Fisica dell'Universit\`a
     di Genova and INFN sezione di Genova, Via Dodecaneso 33, 16146
     Genova, Italy. E-mail: zanghi@ge.infn.it}
}
\date{December 1, 2009}

\newcommand{\be}{\begin{equation}}
\newcommand{\ee}{\end{equation}}

\renewcommand{\Im}{\mathrm{Im}}
\newcommand{\RRR}{\mathbb{R}}
\newcommand{\CCC}{\mathbb{C}}
\newcommand{\scp}[2]{\langle #1|#2 \rangle}
\newcommand{\vQ}{\boldsymbol{Q}}
\newcommand{\vq}{\boldsymbol{q}}
\newcommand{\vj}{\boldsymbol{j}}
\newcommand{\vv}{\boldsymbol{v}}
\newcommand{\NR}{{}^N\RRR}

\begin{document}
\maketitle
\begin{abstract}
Bohmian trajectories have been used for various purposes, including the numerical simulation of the time-dependent Schr\"odinger equation and the visualization of time-dependent wave functions. We review the purpose they were invented for: to serve as the foundation of quantum mechanics, i.e., to explain quantum mechanics in terms of a theory that is free of paradoxes and allows an understanding that is as clear as that of classical mechanics. Indeed, they succeed in serving that purpose in the context of a theory known as Bohmian mechanics, to which this article is an introduction.
\end{abstract}

\section{Bohmian Trajectories}\label{sec:traj}

Let us consider a wave function $\psi_t(q)$ of non-relativistic quantum mechanics, defined on the configuration space $\RRR^{3N}$ of $N$ particles, taking values in the set $\CCC$ of complex numbers, and evolving with time $t$ according to the non-relativistic Schr\"odinger equation,
\be\label{Schr}
i\hbar\frac{\partial\psi_t}{\partial t} = -\sum_{k=1}^N \frac{\hbar^2}{2m_k} \nabla_k^2 \psi_t + V\psi_t\,,
\ee
where $m_k$ is the mass of the $k$-th particle, $\nabla_k=\nabla_{\vq_k}=\bigl(\frac{\partial}{\partial x_{k}},\frac{\partial}{\partial y_{k}}, \frac{\partial}{\partial z_{k}}\bigr)$ is the derivative with respect to the coordinates of the $k$-th particle, and $V:\RRR^{3N}\to \RRR$ is a potential function, for example the Coulomb potential
\be\label{Coulomb}
V(\vq_1,\ldots,\vq_N) = \sum_{1\leq j<k\leq N} \frac{e_je_k}{|\vq_j-\vq_k|}
\ee 
with $e_k$ the charge of the $k$-th particle.

With this wave function there is associated a family of trajectories in configuration space $\RRR^{3N}$, the \emph{Bohmian trajectories}, which are defined to be those trajectories $t\mapsto Q(t)=(\vQ_1(t),\ldots,\vQ_N(t))$ satisfying the equation
\be\label{Bohm}
\frac{d\vQ_k(t)}{dt} = \frac{\hbar}{m_k} \Im \frac{\nabla_k\psi_t}{\psi_t}(Q(t))\,. 
\ee
Put differently, to every wave function $\psi:\RRR^{3N}\to \CCC$ there is associated a vector field $v^\psi$ on configuration space according to
\be\label{vdef}
v^\psi = (\vv_1^\psi,\ldots,\vv_N^\psi)\,, \quad
\vv_k^\psi =\frac{\hbar}{m_k} \Im \frac{\nabla_k\psi}{\psi}\,,
\ee
and \eqref{Bohm} amounts to
\be\label{BohmQv}
\frac{dQ(t)}{dt} = v^{\psi_t}(Q(t))\,.
\ee
Since \eqref{BohmQv} is an ordinary differential equation (ODE) of first order (or rather, a system of $3N$ coupled ODEs of first order), it has, leaving aside the exceptions, a unique solution for every choice of initial configuration $Q(0)$. 

Another way of writing \eqref{BohmQv} is
\be\label{Bohmj}
\frac{dQ(t)}{dt} = \frac{j^{\psi_t}}{|\psi_t|^2}(Q(t))\,,
\ee
where $j^\psi$ is the vector field on $\RRR^{3N}$ usually called the \emph{probability current} associated with the wave function $\psi$,
\be\label{jdef}
j^\psi = (\vj_1^\psi,\ldots,\vj_N^\psi)\,, \quad
\vj_k^\psi =\frac{\hbar}{m_k} \Im (\psi^*\nabla_k\psi)\,.
\ee

\section{Bohmian Mechanics}
\label{sec:BM}

The theory that uses Bohmian trajectories as the foundation of quantum mechanics is known as Bohmian mechanics; it arises if we take a Bohmian trajectory seriously. Namely, Bohmian mechanics claims that in our world, electrons and other elementary particles have precise positions $\vQ_k(t)\in\RRR^3$ at every time $t$ that move according to \eqref{Bohm}. That is, for a certain Bohmian trajectory $t\mapsto Q(t)$ in configuration space, it claims that $Q(t)=(\vQ_1(t),\ldots,\vQ_N(t))$ is the configuration of particle positions in our world at time $t$. 

This picture is in contrast with the orthodox view of quantum mechanics, according to which quantum particles do not have precise positions, but are regarded as ``delocalized'' to the extent to which the wave function $\psi_t$ is spread out. It is also in contrast with another picture of the Bohmian trajectories that one often has in mind when using Bohmian trajectories for numerical purposes: the hydrodynamic picture. According to the latter, all the Bohmian trajectories associated with a given wave function (but corresponding to different $Q(0)$) are on an equal footing, none is more real than the others, they are all regarded as flow lines in analogy to the flow lines of a classical fluid. In Bohmian mechanics, however, only one of the Bohmian trajectories corresponds to reality, and all the other ones are no more than mathematical curves, representing possible alternative histories that could have occurred if the initial configuration of our world had been different, but did not occur. 

As a consequence, talk of probability makes immediate sense in Bohmian mechanics but not in the hydrodynamic picture: In Bohmian mechanics, with only one trajectory realized, that trajectory may be random. In the hydrodynamic picture, with all trajectories equally real, it is not clear what a probability distribution over the trajectories could be the probability \emph{of}, and what it could mean to say that a trajectory is random.

Bohmian mechanics was first proposed by Louis de Broglie (1892--1987) in the 1920s \cite{deB28}; it is named after David Bohm (1917--1992), who was the first to realize that this theory provides a foundation for quantum mechanics \cite{Bohm52}: The inhabitants of a typical Bohmian world would, as a consequence of the equations of Bohmian mechanics, observe exactly the probabilities predicted by the quantum formalism. 

To understand how this comes about requires a rather subtle ``quantum equilibrium'' analysis \cite{qe} that is beyond the scope of this paper. An important element of the analysis is however rather simple. It is the property of {\it equivariance}, expressing the compatibility between  the   evolution of the wave function given by Schr\"odinger's equation and the evolution of the actual configuration given by the {\it guiding equation} \eqref{Bohmj}. This property will be discussed in the next section.

The upshot of the quantum equilibrium analysis is the justification of the {\it probability postulate} for Bohmian mechanics, that the configuration $Q$ of a system with wave function $\psi=\psi(q)$ is random with probability density $|\psi(q)|^2$. Bohmian mechanics, with the probability postulate, is \emph{empirically equivalent} to standard quantum mechanics. We will return to this point later and explain how this follows from the equations.

\section{Equivariance}

Equivariance amounts to the following assertion: If $Q(0)$ is random with probability density given by $|\psi_0|^2$, then  $Q(t)$ is also random, with probability density given by $|\psi_t|^2$. 

This is  easy to see: Let $\rho_t$ be the probability density of $Q(t)$. Then $\rho_t$ evolves according to the continuity equation
\be\label{cont}
\frac{\partial\rho_t}{\partial t} = -\mathrm{div} (\rho_t v^{\psi_t})\,,
\ee
whose right hand side is short for 
\[
-\sum_{k=1}^N \nabla_k \cdot (\rho_t \vv_k^{\psi_t})\,,
\]
where $\cdot$ denotes the dot product of two vectors in $\RRR^3$. On the other hand, from the Schr\"odinger equation \eqref{Schr}
\be
\frac{\partial}{\partial t}\bigl( \psi_t^* \psi_t \bigr) =
-\mathrm{div}\, j^{\psi_t}\,,
\ee
which is the same equation as \eqref{cont} with $\rho_t$ replaced by $|\psi_t|^2$. Thus, if $\rho_0=|\psi_0|^2$ then $\rho_t=|\psi_t|^2$ at any time $t$.

\section{The Quantum Potential}

A particle trajectory $\vQ(t)$ can always be written in the form of Newton's law:
\be
m \frac{d^2\vQ(t)}{dt^2} = \text{force}(t)\,,
\ee
where $m$ is the mass of the particle; after all, the right hand side can simply be so chosen as to make this equation true. It is sometimes useful to do this for the trajectories of the Bohmian particles; we find, by taking the time derivative of \eqref{Bohm} and after some calculation:
\be\label{Q2}
m_k\frac{d^2\vQ_k(t)}{dt^2} = -\nabla_k (V+V_{qu}^{\psi_t})(Q(t))\,,
\ee
where $V_{qu}^\psi$ is a function called the \emph{quantum potential},
\be\label{Vqudef}
V_{qu}^\psi = -\sum_{j=1}^N \frac{\hbar^2}{2m_j} \frac{\nabla_j^2|\psi|}{|\psi|}\,.
\ee
For comparison, a classical particle would move according to
\be\label{classical}
m_k\frac{d^2\vQ_k(t)}{dt^2} = -\nabla_k V(Q(t))\,.
\ee
That is, a Bohmian trajectory is also a solution to classical mechanics if we add a suitable time-dependent term $V_{qu}^{\psi_t}$ to the potential function $V$. One particular application of the quantum potential arises in the study of the classical limit of quantum mechanics \cite{ADGZ02}: As we see from \eqref{Q2} and \eqref{classical}, the regime in which Bohmian trajectories agree with classical trajectories is characterized by the condition that the gradient of the quantum potential vanishes, or at least, for approximate agreement, is small.

Several things have been, or may be, puzzling about the quantum potential. In David Bohm's 1952 article \cite{Bohm52} on Bohmian mechanics (of course, Bohm did not refer to this theory as ``Bohmian mechanics''), he presented \eqref{Q2} together with \eqref{Vqudef}, rather than \eqref{Bohm}, as the basic equation of motion. That created a sense of mystery because it does not appear natural to postulate the existence of an additional potential without specifying which physical object causes this potential, and how and why. Moreover, since the formula \eqref{Vqudef} is neither obvious nor natural, it created the impression that Bohmian mechanics was a contrived, artificial theory. These unnecessary difficulties vanish when we regard \eqref{Bohm} as the equation of motion because then we do not just add another term to \eqref{classical} but replace it instead with an equation that is altogether different but equally simple. Furthermore, \eqref{Q2} is mathematically not equivalent to \eqref{Bohm}: while every solution of \eqref{Bohm} is a solution of \eqref{Q2}, the converse is not true, as \eqref{Q2} is a second-order equation that provides a solution for every choice of initial positions and velocities, including choices for which the initial velocities fail to be related to the initial positions in accordance with \eqref{Bohm}. Bohm introduced, in order to exclude these further solutions of \eqref{Q2}, a constraint condition on the possible velocities---and the condition was \eqref{Bohm}! In fact, it follows from the fact that \eqref{Q2} can be obtained from \eqref{Bohm} that if a solution of the second-order equation \eqref{Q2} has initial velocities satisfying \eqref{Bohm} then also the velocities at any other time will satisfy \eqref{Bohm}. But then \eqref{Bohm} is satisfied at all times, and instead of calling it a constraint we can simply call it the equation of motion. To call it a constraint just creates another unnecessary mystery: why should nature impose such a constraint? Would it not be simpler to have only equation \eqref{Q2}? Not, of course, if the alternative is to have only equation \eqref{Bohm}.

\section{Connection with Numerical Methods}
\label{sec:numerical}

In our definition \eqref{Bohm} of the Bohmian trajectories, we used $\psi_t$ for every $t$, supposing that the Schr\"odinger equation has been solved already. It may thus seem surprising that the Bohmian trajectories can be used for solving the Schr\"odinger equation (up to a global phase factor). But the second-order equation \eqref{Q2} suggests how that can be done, roughly as follows \cite{Lopreore}: 

Suppose we know the initial wave function $\psi_0$. Choose an ensemble of points (say $Q^{(1)},\ldots,Q^{(n)}$, with very large $n$) in configuration space; note the difference between $N$ points in physical space $\RRR^3$ (which is one configuration) and $n$ points in configuration space $\RRR^{3N}$ (which correspond to $nN$ points in $\RRR^3$). Choose the ensemble so that its distribution density in configuration space is, with sufficient accuracy, $\rho_0=|\psi_0|^2$. For each $Q^{(i)}$, determine $v^{\psi_0}(Q^{(i)})$, the $3N$-vector of velocities, from the velocity law \eqref{vdef}. For each $i$, solve the second-order equation of motion \eqref{Q2} with initial positions as in $Q^{(i)}$, initial velocities as in $v^{\psi_0}(Q^{(i)})$, and the quantum potential as defined in \eqref{Vqudef} but with $|\psi_t|$ replaced by $\sqrt{\rho_t}$, where $\rho_t$ is the density in configuration space of the ensemble $Q^{(1)}(t),\ldots,Q^{(n)}(t)$. That is, for every time step $t \to t+\delta t$, determine the quantum potential at time $t$ from the density $\rho_t$ of the ensemble points according to
\be\label{Vqurho}
V_{qu}(t) = -\sum_{j=1}^N \frac{\hbar^2}{2m_j} \frac{\nabla_j^2\sqrt{\rho_t}}{\sqrt{\rho_t}}\,,
\ee
then use \eqref{Q2} to propagate $Q^{(i)}(t)$ and $\frac{dQ^{(i)}}{dt}(t)$ by one time step and obtain $Q^{(i)}(t+\delta t)$ and $\frac{dQ^{(i)}}{dt}(t+\delta t)$. 

Equivariance, together with the fact that \eqref{Q2} follows from the Schr\"odinger equation \eqref{Schr} and the equation of motion \eqref{Bohm}, implies that this method would yield exactly the family of Bohmian trajectories if we used infinitely many sample points with an initial distribution given \emph{exactly} by $|\psi_0|^2$, if the time step $\delta t$ were infinitesimal, and if no numerical error were involved in solving \eqref{Q2}. With finite $n$ and finite $\delta t$, we may obtain an approximation to the family of Bohmian trajectories. 

Once we have the trajectories, we can (more or less) recover the wave function $\psi_t(q)$ up to a time-dependent phase factor as follows. Note that the right hand side of the equation of motion \eqref{Bohm} is proportional to the $\vq_k$-derivative of the phase of the wave function; i.e., if we write 
\be
\psi(q) = |\psi(q)| \, e^{iS(q)/\hbar}
\ee
with a real-valued function $S$ then
\be\
\vv_k^{\psi}(q) = \frac{1}{m_k} \nabla_k S\,.
\ee
Now, if we know all Bohmian trajectories then we can read off the $3N$-velocity $v(q,t)=(\vv_1(q,t),\ldots,\vv_N(q,t))$ of the trajectory that passes through $q\in\RRR^{3N}$ at time $t$, solve
\be\label{vS}
\vv_k(q,t) = \frac{1}{m_k} \nabla_k S(q,t)
\ee
for $S(q,t)$, and finally set
\be
\psi_t(q) = \sqrt{\rho_t(q)} e^{iS(q,t)/\hbar}\,.
\ee
Note that \eqref{vS} determines the function $q\mapsto S(q,t)$ up to a real constant $\theta(t)$; i.e., any $S(q,t)+\theta(t)$ is another solution of \eqref{vS} and, conversely, if $S_1(q,t)$ and $S_2(q,t)$ are two solutions of \eqref{vS} then $S_2(q,t)-S_1(q,t)$ is (real and) independent of $q$ and can thus be called $\theta(t)$. As a consequence, $\psi_t(q)$ has been determined up to a (global, i.e., $q$-independent) phase factor $e^{i\theta(t)}$.

\section{The Quantum Potential Again}

The previous section has illustrated how the second-order equation \eqref{Q2} and the concept of the quantum potential can be useful even if we regard the first-order equation \eqref{Bohm} as the fundamental equation of motion. But the algorithm outlined there leads to another puzzle about the quantum potential: It involves a picture in which an ensemble of points in configuration space, with density $\rho_t$, leads to a quantum potential via \eqref{Vqurho}, which in turn acts on every trajectory. This may suggest that this ensemble of points in configuration space is the physical \emph{cause} of some kind of real field, the quantum potential, which in turn is the physical \emph{cause} of the shape of the individual trajectory, in particular of its deviation from a classical trajectory. This picture, however, requires that all Bohmian trajectories be physically real, in agreement with the hydrodynamic picture mentioned before, but in conflict with Bohmian mechanics as described before, the theory asserting that only one of the trajectories is real while the others are merely hypothetical. But how could merely hypothetical trajectories push the actual particles around? They cannot. Bohmian mechanics is incompatible with the picture that the density of trajectories causes a quantum potential that pushes in turn every trajectory.

\section{Wave--Particle Duality}

So what picture arises instead from Bohmian mechanics? The object that influences the motion of the one actual configuration is the wave function. Note, however, that we should not assume that the configuration would ``normally'' move along a classical trajectory unless some physical agent (be it other trajectories, the quantum potential, or the wave function) pushed it off to another trajectory; rather, the classical equation of motion \eqref{classical} is replaced by a new equation of motion \eqref{Bohm}, and this new equation does not talk about forces, about pushing, or about causes, but merely defines the trajectory in terms of the wave function. By virtue of the very purpose that it was designed for, the numerical algorithm above avoids referring to the wave function; after all, it is an algorithm for \emph{finding} the wave function. However, for a theory such as Bohmian mechanics (that is, for a proposal as to how nature might work), it is acceptable to suppose that nature solves the Schr\"odinger equation independently of trajectories, and then lets the one trajectory depend on the wave function. This is the picture that arises if we insist that only one trajectory is real.

As a consequence, we need to take the wave function seriously as a physical object. Put differently, in a world governed by Bohmian mechanics, there is a wave--particle duality in the literal sense: there is a wave ($\psi$ on $\RRR^{3N}$), and there are particles (at $\vQ_1,\ldots,\vQ_N$). The wave evolves according to the Schr\"odinger equation \eqref{Schr}, and the particles move in a way that depends on the wave, namely according to \eqref{Bohm}. Put differently, the wave guides, or pilots, the particles; that is why this theory has also been called the \emph{pilot-wave theory}.

\section{The Phase Function $S(q,t)$}

Above in Section~\ref{sec:numerical}, we have made use of writing the wave function in terms of its modulus, often denoted $R(q,t)=|\psi(q,t)|$, and phase function, often denoted $S(q,t)/\hbar$,
\be\label{psiRS}
\psi(q,t) = R(q,t) \, e^{iS(q,t)/\hbar}\,.
\ee
There are some difficulties with this decomposition, though, and maybe this is a good place to make the reader aware of them. 

The first difficulty is that the value of $S(q,t)$ is not uniquely determined by \eqref{psiRS}, but only up to addition of an integer multiple of $2\pi\hbar$. Of course, we can simply choose one of the possible values, but it is not always possible to stick with that choice; more precisely, it is not always possible to choose $S$ as a continuous function, even though $\psi$ is continuous. An example can be found among the eigenstates of the hydrogen atom, which are known to factorize in spherical coordinates into a function of the radial coordinate $r$, a function of the colatitude $\theta$, and a function of the azimuth $\varphi$; 
what matters here is that the last factor is $e^{im\varphi}$ with integer $m$, which contributes a summand $m\varphi\hbar$ to the $S$ function. For $m\neq 0$, this $S$ function is discontinuous, as it jumps from $m2\pi\hbar$ to $0$ at $\varphi=2\pi$, while $\psi$ is continuous. (In fact, every choice of the $S$ function will be discontinuous, since for any fixed $r$ and $\theta$ such that $\psi(r,\theta,\varphi)\neq 0$,
\be
\frac{\partial S}{\partial \varphi}=\hbar\,\Im\Bigl(\frac{1}{\psi}\frac{\partial\psi}{\partial \varphi}\Bigr) = m\hbar\,,
\ee
so $S$ has to grow linearly with $\varphi$.)

Note that, at the discontinuity, the $S$ function cannot jump by an arbitrary amount, but only by an integer multiple of $2\pi\hbar$, and that $S$ is not defined where $\psi=0$. As a consequence, the correspondence between a complex-valued function $\psi$ and the two real-valued functions $R$ and $S$ is a bit complicated. And as a consequence of \emph{that}, the usual pair of real equations for $R$ and $S$ that can be obtained from the Schr\"odinger equation,
\begin{align}
\frac{\partial R^2}{\partial t} &= -\sum_{k=1}^N \nabla_k\cdot \biggl(R^2 \frac{\nabla_k S}{m_k}\biggr)\label{Req}\\
\frac{\partial S}{\partial t} &= \sum_{k=1}^N \biggl(\frac{\hbar^2}{2m_k}\frac{\nabla_k^2 R}{R}-\frac{(\nabla_k S)^2}{2m_k}\biggr) - V \,,\label{Seq}
\end{align}
are actually \emph{not} equivalent to Schr\"odinger's equation. Explicitly, if we started from \eqref{Req} and \eqref{Seq} then we would have no reason to allow $S$ to be undefined where $R=0$; we would have no reason to expect discontinuities in $S$; and if we allowed discontinuities then we would have no reason to demand that the jump height is an integer multiple of $2\pi\hbar$.

\section{Predictions and the Quantum Formalism}

Consider a hypothetical world governed by Bohmian mechanics, and let us call this a Bohmian world. We have mentioned already in Section~\ref{sec:BM} that the inhabitants of a Bohmian world would observe exactly the probabilities predicted by the quantum formalism. In this section, we outline why this is so.

Consider an experiment carried out by an observer. In a Bohmian world, of course, also observers and their apparatuses (the detectors, cameras, photographs, display screens, meter pointers, etc.) consist of particles governed by the equations of Bohmian mechanics. For this reason, let us for a moment consider the $N$-particle system formed by both, the object of the experiment and the apparatus. Let us write the configuration of our system as $Q=(X,Y)\in\RRR^{3N}$ with $X\in\RRR^{3K}$ the configuration of the object and $Y\in\RRR^{3L}$ the configuration of the apparatus. The number $N=K+L$ will be huge because $L$ is, in fact usually $L>10^{23}$, as the apparatus is a macroscopic system; $K$, in contrast, may be just 1. Correspondingly, we write the wave function $\Psi$ of this $N$-particle system as
\be
\Psi(q) = \Psi(x,y)\,.
\ee
Suppose that, at the time $t_0$ at which the experiment begins, the wave function factorizes,
\be\label{Psit0}
\Psi_{t_0}(x,y) = \psi(x)\, \phi(y)
\ee
with $\psi$ the wave function of the object at time $t_0$ and $\phi$ the initial state (``ready state'') of the apparatus, $\scp{\psi}{\psi}=1=\scp{\phi}{\phi}$. (Actually, the symmetrization postulate implies that $\Psi_{t_0}$ cannot factorize in this way if both the object and the apparatus contain particles of the same species, say, if both contain electrons. A treatment that takes the symmetrization postulate into account leads to the same conclusions but is harder to follow, and we prefer to simplify the discussion.) 

Suppose the experiment is over at time $t_1$. The wave function at that time is, of course, given by
\be\label{Psit1}
\Psi_{t_1}= e^{-iH(t_1-t_0)/\hbar} \Psi_{t_0}
\ee
with $H$ the Hamiltonian of the $N$-particle system. (We are assuming, for simplicity, that the system is isolated during the experiment; this is not a big assumption since we could make the $N$-particle system as large as we want, even comprising the entire universe.)

Suppose further that for certain wave functions $\psi_\alpha(x)$ that the object might have, the apparatus will yield a predictable result $r_\alpha$. More precisely, suppose that if $\psi=\psi_\alpha$ in \eqref{Psit0} then the final wave function $\Psi_{t_1}$ is concentrated on the set $S_\alpha$ of those $y$-configurations in which the apparatus' pointer points to the value $r_\alpha$,
\be
\int\limits_{\RRR^{3K}} dx \int\limits_{S_\alpha} dy\, |\Psi_{t_1}(x,y)|^2 = 1  \,.
\ee
(For example, standard quantum mechanics asserts that this is the case when the experiment is a ``quantum measurement of the observable with operator $A$'' and $\psi_\alpha$ is an eigenfunction of $A$ with eigenvalue $r_\alpha$. In general, an analysis of the experiment shows whether this is the case.)

Let us write $\Psi^{(\alpha)}$ for $\Psi_{t_1}$ arising from $\psi=\psi_\alpha$, i.e.,
\be
\Psi^{(\alpha)} = e^{-iH(t_1-t_0)/\hbar} (\psi_{\alpha} \phi)\,.
\ee
It now follows from the linearity of the Schr\"odinger equation that if the wave function of the object is a (non-trivial) linear combination of the $\psi_\alpha$,
\be
\psi=\sum_\alpha c_\alpha \, \psi_\alpha
\ee
then
\be\label{after}
\Psi_{t_1} = \sum_\alpha c_\alpha \, \Psi^{(\alpha)}\,,
\ee
which is a (non-trivial) superposition of different wave functions that correspond to different outcomes (and macroscopically different orientations of the pointer). It is known as the \emph{measurement problem of quantum mechanics} that this wave function, the wave function of the object and the apparatus together after the experiment as determined by the Schr\"odinger equation, does not single out one of the $r_\alpha$s as the actual outcome of the experiment.

Since Bohmian mechanics assumes the Schr\"odinger equation, \eqref{after} is the correct wave function in Bohmian mechanics. Moreover, the configuration $Q(t_1)=(X(t_1),Y(t_1))$ is, by the probability postulate and equivariance, random with probability density $|\Psi_{t_1}|^2$. As a consequence, $Y_{t_1}$ lies in the set $S_\alpha$ with probability
\be
\int\limits_{\RRR^{3K}}dx\int\limits_{S_\alpha} dy \, |\Psi_{t_1}(x,y)|^2 =
\int\limits_{\RRR^{3K}}dx\int\limits_{S_\alpha} dy \, |c_\alpha|^2 \, |\Psi^{(\alpha)}(x,y)|^2 = |c_\alpha|^2
\ee
because $\Psi^{(\beta)}(x,y)=0$ for $y\in S_\alpha$ and $\beta\neq \alpha$. But that $Y_{t_1}$ lies in the set $S_\alpha$ means that the pointer is pointing to the value $r_\alpha$. Thus, in Bohmian mechanics the apparatus (consisting of Bohmian particles) does point to a certain value, and the value is always one of the $r_\alpha$'s (the same values as provided by the quantum formalism), and the value is random, and the probability it is $r_\alpha$ is $|c_\alpha|^2$ (the same probability as provided by the quantum formalism).

\section{The Generalized Quantum Formalism}

The above example illustrates why Bohmian mechanics predicts the same probabilities for the results of quantum measurements as the standard quantum formalism. Let us see what we obtain if we drop the assumption that for $\psi=\psi_\alpha$ the experiment yields a predictable result $r_\alpha$. It is still true, then, that if the configuration $Y_{t_1}$ of the apparatus lies in the set $S_\alpha\subset \RRR^{3L}$ then the pointer points to the value $r_\alpha$, the result of the experiment. The probability that that happens is
\be
p_\alpha:=\int\limits_{\RRR^{3K}}dx\int\limits_{S_\alpha} dy \, |\Psi_{t_1}(x,y)|^2\,.
\ee
A calculation then shows that there is a positive self-adjoint (and uniquely determined) operator $E_\alpha$ such that
\be\label{palpha}
p_\alpha = \scp{\psi_{t_0}}{E_\alpha|\psi_{t_0}}\,.
\ee
Indeed, 
\be
E_\alpha = \scp{\phi}{e^{iH(t_1-t_0)/\hbar}(I\otimes 1_{S_\alpha})e^{-iH(t_1-t_0)/\hbar}|\phi}_y\,,
\ee
where $\scp{\cdot}{\cdot}_y$ is the partial scalar product taken only over $y$ but not over $x$, $I$ is the identity operator (in this case, on the $x$-Hilbert space), and $1_{S_\alpha}$ is the operator that multiplies by the characteristic function of the set $S_\alpha$. If the sum of the $p_\alpha$ is 1 for every $\psi_{t_0}$ (which is the case if we can neglect the possibility that the experiment fails to yield any result, and if we have introduced sufficiently many sets $S_\alpha$ so as to cover all possible results $r_\alpha$) then
\be\label{sumEalpha}
\sum_\alpha E_\alpha = I\,.
\ee

A family of positive operators $\{E_\alpha\}$ obeying \eqref{sumEalpha} is called a \emph{positive-operator-valued measure (POVM)}, and the rule that the probability of the result $r_\alpha$ is given by \eqref{palpha} is part of the \emph{generalized quantum formalism}. In case each $E_\alpha$ is a projection operator, we obtain back the usual rules of quantum measurement, as the operator
\be
A=\sum_\alpha r_\alpha \, E_\alpha
\ee
is self-adjoint with eigenvalues $r_\alpha$, and $E_\alpha$ is the projection to the eigenspace of $A$ with eigenvalue $r_\alpha$.

\section{What is Unsatisfactory About Standard Quantum Mechanics}

Many physicists, beginning with Einstein and Schr\"odinger and including the authors, have felt that standard quantum mechanics is not satisfactory as a physical theory. This is because the axioms of standard quantum mechanics concern the results an observer will obtain if he performs a certain experiment. We think that a fundamental physical theory should not be formulated in terms of concepts like ``observer'' or ``experiment,'' as these concepts are very vague and certainly do not seem fundamental. Is a cat an observer? A computer? Were there any experiments before life existed on Earth? Instead, a fundamental physical theory should be formulated in terms of rather simple physical objects in space and time like fields, particles, or perhaps strings. Bohmian mechanics is a beautiful example of a theory that is satisfactory as a fundamental physical theory.

It is a frequent misunderstanding that the main problem that the critics of standard quantum mechanics have is that it is different from classical mechanics. A fundamental physical theory can very well be different from classical mechanics, but we should demand that it be as \emph{clear} as classical mechanics. We think that, for it to make clear sense as physics, it must describe matter moving in space. Standard quantum mechanics does not do that, but Bohmian mechanics does, as it describes the motion of point particles. And, indeed, Bohmian mechanics is different from classical mechanics in many crucial respects. Another frequent misunderstanding is that the goal of Bohmian mechanics is to return as much as possible to classical mechanics. The circumstance that it uses point particles and is deterministic does not mean that we are dogmatically committed to point particles or determinism; also indeterministic theories with another ontology may very well be satisfactory. It just so happens that the simplest satisfactory version of quantum mechanics, Bohmian mechanics, involves point particles and determinism.

It is also a misunderstanding to think that the goal of Bohmian mechanics was to \emph{derive} the Schr\"odinger equation, or to \emph{replace} it with something else. Of course, a physical theory may involve new postulates and introduce new equations, and we see no problem with the Schr\"odinger equation. Rather, the goal is to replace the \emph{measurement postulate} of standard quantum mechanics with postulates that refer to electrons and nuclei instead of observers, axioms from which the measurement rules can be derived as theorems.

After these more philosophical themes, let us finally turn to two more technical topics, how to incorporate spin and identical particles into Bohmian mechanics.

\section{Spin}

In order to treat particles with spin, almost no change in the defining equations of Bohmian mechanics is necessary. Recall that the wave function of a spin-$\tfrac12$ particle can be regarded as a function $\psi:\RRR^3\to\CCC^2$, and for $N$ such particles as $\psi:\RRR^{3N}\to\CCC^{2^N}$. That is, $\psi$ is now a multi-component (or vector-valued) function. We keep the form \eqref{Bohmj}, i.e., $dQ/dt=j^\psi/|\psi|^2$, of the equation of motion, with the appropriate expressions for $j^\psi$ and $|\psi|^2$ in terms of a multi-component function $\psi$; viz., we postulate, as the equation of motion, \cite{Bell66}
\be\label{Bohmspin}
\frac{d\vQ_k(t)}{dt} = \frac{(\hbar/m_k) \Im(\psi_t^\dagger\nabla_k\psi_t)}{\psi_t^\dagger\psi_t}(Q(t))\,,
\ee
where
\be\label{dagger}
\phi^\dagger\psi = \sum_{s=1}^{2^N} \phi_s^* \psi_s
\ee
is the scalar product in $\CCC^{2^N}$. Particles with spin other than $\tfrac12$ can be treated in a similar way. 

If we introduce an external magnetic field $\boldsymbol{B}$, the Schr\"odinger equation needs to be modified appropriately, i.e., replaced (as usual) with the Pauli equation
\be\label{Pauli}
i\hbar\frac{\partial\psi_t}{\partial t} = -\sum_{k=1}^N \frac{\hbar^2}{2m_k} \bigl(\nabla_k-ie_k\boldsymbol{A}(\vq_k)\bigr)^2 \psi_t + \sum_{k=1}^N \mu_k \boldsymbol{B}(\vq_k)\cdot \boldsymbol{\sigma}_k\psi_t+ V\psi_t\,,
\ee
where $\boldsymbol{A}$ is the vector potential, $e_k$ and $\mu_k$ are the charge and the magnetic moment of the $k$-th particle, $\boldsymbol{\sigma}=(\sigma_x,\sigma_y,\sigma_z)$ is the vector consisting of the three Pauli spin matrices, and $\boldsymbol{\sigma}_k$ acts on the spin index $s_k\in\{+1,-1\}$ that is associated with the $k$-th particle when we regard the spin component index $s$ in \eqref{dagger} as a multi-index, $s=(s_1,\ldots,s_N)$.

In this theory, since particles are not literally spinning (i.e., not rotating), the word ``spin'' is an anachronism like the pre-Copernican word \emph{sunrise}. What may be more surprising about the formulation of Bohmian mechanics for particles with spin is that we did not have to introduce further (``hidden'') variables, on the same footing as the positions $\vQ_k(t)$, and that there is no direction of space in which the ``spin vector'' is ``actually'' pointing. In particular, a Stern--Gerlach experiment does not \emph{measure} the value of an additional spin variable. How can that be? In a Stern--Gerlach experiment, we arrange an external magnetic field for the particle to pass through, and measure its position afterwards; depending on where the particle was detected, we say that we obtained the result ``up'' or ``down.'' Using equivariance, which holds for \eqref{Bohmspin} just as it does for \eqref{Bohmj}, one easily sees that this experiment yields, in Bohmian mechanics, a random result that has the same probability distribution as predicted by standard quantum mechanics. Looking at the experiment in this way, what is happening is completely clear and it is certainly not necessary to postulate that the particle has an actual spin vector, before the experiment or after it.

\section{The Symmetrization Postulate}

Our description of Bohmian trajectories in Section~\ref{sec:traj} did not include the proper treatment of identical particles. The \emph{symmetrization postulate} of quantum mechanics asserts that if the variables $\vq_i$ and $\vq_j\in\RRR^3$ in the wave function $\psi$ refer to two identical particles (i.e., two particles of the same species, such as, e.g., two electrons) then $\psi$ is either \emph{symmetric} in $\vq_i$ and $\vq_j$,
\be\label{sym}
\psi(\ldots \vq_i \ldots \vq_j \ldots) = \psi( \ldots \vq_j \ldots \vq_i \ldots)\,,
\ee
if the species is \emph{bosonic}, or \emph{anti-symmetric} in $\vq_i$ and $\vq_j$,
\be\label{anti}
\psi(\ldots \vq_i \ldots \vq_j \ldots) =- \psi(\ldots \vq_j \ldots \vq_i \ldots)\,,
\ee
if the species is \emph{fermionic}. In equations \eqref{sym} and \eqref{anti} it is understood that all other variables remain unchanged, only the variables $\vq_i$ and $\vq_j$ get interchanged. The \emph{spin-statistics rule} asserts that every species with integer spin ($0,1,2,\ldots$) is bosonic and every species with half-odd spin ($\tfrac12, \tfrac32, \ldots$) is fermionic.

In Bohmian mechanics for identical particles, one uses the same type of wave function as in ordinary quantum mechanics, and the same equation of motion (and Schr\"odinger equation) as in Bohmian mechanics for distinguishable particles. That is, we include the symmetrization postulate among the postulates of Bohmian mechanics.

It is a traditional claim in textbooks on quantum mechanics that the
lack of precise trajectories in orthodox quantum mechanics is the
reason for the symmetrization postulate. If the particles had
trajectories, it is suggested, then they would automatically be distinguishable. 
From Bohmian mechanics with the symmetrization postulate we see that this suggestion is incorrect.

On the contrary, the Bohmian trajectories actually \emph{enhance} our understanding of the symmetrization postulate. We start from the following observation: 
\begin{align}\label{Bohmsym}
\text{If }&(\ldots \vQ_i(0) \ldots \vQ_j(0) \ldots) \text{ evolves to } (\ldots \vQ_i(t) \ldots \vQ_j(t) \ldots)\nonumber\\
\text{ then }&(\ldots \vQ_j(0) \ldots \vQ_i(0) \ldots) \text{ evolves to } (\ldots \vQ_j(t) \ldots \vQ_i(t) \ldots)\,.
\end{align}
In other words, it is unnecessary to specify the \emph{labelling} of the particles. That seems very appropriate as the labelling is unphysical. The fact \eqref{Bohmsym} follows from the symmetry of the velocity vector field,
\begin{align}
\vv_i(\ldots \vq_i \ldots \vq_j \ldots ) &= \vv_j(\ldots \vq_j \ldots \vq_i \ldots)\,,\\
\vv_k(\ldots \vq_i \ldots \vq_j \ldots ) &= \vv_k(\ldots \vq_i \ldots \vq_j \ldots)
\quad\text{for }i\neq k \neq j\,,
\end{align}
which can easily be checked from the formula \eqref{vdef} for $\vv_i$ together with \eqref{sym} or \eqref{anti}. For another way of putting the fact \eqref{Bohmsym}, let us consider a system of $N$ identical particles. The natural configuration space is
\be
\NR^3 = \{Q\subset \RRR^3: \#Q=N\}\,,
\ee
the set of all $N$-element subsets of $\RRR^3$. While an element of $\RRR^{3N}$ is an \emph{ordered} configuration $(\vQ_1,\ldots,\vQ_N)$, an element of $\NR^3$ is an \emph{unordered} configuration $\{\vQ_1,\ldots,\vQ_N\}$. Since the labels need not be specified, any point $Q(0)\in\NR^3$ as initial condition will uniquely define a curve $t\mapsto Q(t)\in\NR^3$. So for symmetric or anti-symmetric wave functions, Bohmian mechanics works on the natural configuration space of identical particles. This fact can be regarded as something like an explanation, or derivation, of the symmetrization postulate. For a deeper discussion see \cite{DGTZ07}.

\section{Further Reading}

Concerning the extension of Bohmian mechanics to quantum field theory, see \cite{DGTZ04,Struyve}. Concerning the extension of Bohmian mechanics to relativistic space-time, see the review article \cite{Tum07} and references therein. Concerning quantum nonlocality, see Bell's book \cite{Bell87b}, which we highly recommend also about foundations of quantum mechanics in general.

\end{document}